\documentstyle{l-aa}
\input{epsf}

\begin{document}

  \thesaurus{02.          
              (08.14.1;   
               02.04.1;   
               02.13.1)   
            }
\title{
Electron conduction along quantizing magnetic fields
in neutron star crusts}
\subtitle{
II. Practical formulae
}

\author{A.Y.\,Potekhin$^{1,2~}$\thanks{
E-mail: palex@astro.ioffe.rssi.ru} 
\and D.G.\,Yakovlev$^1$}
\institute{$^1$A.F.\,Ioffe Physical-Technical Institute,  
           194021, St-Petersburg, Russia \\
           $^2$Nordita, Blegdamsvej 17, DK-2100 Copenhagen \O, Denmark}
\offprints{A.Y.\,Potekhin (Ioffe Institute)}

\date{Received 26 October 1995 / Accepted 17 February 1996}

\maketitle

\markboth{A.Y.\,Potekhin \& D.G.\,Yakovlev: 
Electron conduction along quantizing magnetic fields. II}{}
\newcommand{\ep}{\varepsilon}                  
\newcommand{\vect}[1]{\mbox{\boldmath $#1$}}   
\newcommand{\kB}{\mbox{$k_{\rm B}$}}           

\begin{abstract}
We derive practical expressions
for a rapid and accurate evaluation of
electric and thermal conductivities
and thermopower of degenerate relativistic electrons
along quantizing magnetic fields in outer neutron star crusts.
We consider the electron Coulomb scattering on ions in liquid matter,
as well as on high-temperature phonons or charged impurities
in solid matter. We propose also a reasonable semi-quantitative
treatment of low-temperature phonons.
The transport properties are expressed through
the energy dependent effective electron relaxation time
averaged over electron energies. We calculate
this relaxation time, using
the theoretical formalism of the
previous work, obtain accurate fitting expressions,
and propose an efficient energy
averaging procedure. We create a computer
code which calculates the longitudinal transport properties
of degenerate electrons in strong magnetic fields for any
parameters of dense stellar matter of practical interest.
We analyse quantum oscillations of the transport coefficients
versus density at various temperatures and magnetic
fields.

\keywords{stars: neutron -- dense matter -- 
magnetic fields}

\end{abstract}

\section{Introduction}                                        
\label{sect1}
Transport properties of neutron star
crusts are important for studies of thermal
evolution (cooling) of neutron stars,
evolution of their magnetic fields, etc.
For instance, the heat in the outer crusts of
magnetized neutron stars
is mainly transported by electrons
along the magnetic fields.
As a rule, the electrons in the crust are 
strongly degenerate, and they may be
relativistic. The main electron scattering mechanisms are
the Coulomb scattering on ions in the liquid phase, and
the scattering on phonons or charged
impurities in the solid phase;
the magnetic field can quantize electron motion.

In the present work we study the
electron transport properties in the outer
crusts of neutron stars along quantizing
magnetic fields. The problem has been
considered earlier by several authors.
The first were Canuto and his colleagues in 1970s; the
results were summarized by Canuto \& Ventura (1977).
An adequate kinetic equation for the electron
distribution function was proposed by Yakovlev (1980),
and used for calculating the
transport coefficients
by Yakovlev (1980, 1984), Hernquist (1984), Van Riper (1988)
and Schaaf (1988). Although much work has been done,
these studies are not fully complete.
First, the approach based on the distribution function
formalism is not invariant: the results depend slightly on the choice
of the electron wave functions
(due to the spin degeneracy)
in the magnetic field.
Second, the results are not easy for practical use.
Third, the previous studies neglected
the Debye--Waller reduction of electron-phonon scattering rate
in solid magnetized matter,
although the importance of the Debye--Waller factor
was demonstrated by Itoh et al.\ (1984, 1993) for the
non-magnetic case.

Recently Potekhin (1995, hereafter Paper I) has developed an
invariant formalism of the longitudinal electron conduction
problem based on the spin polarization
density matrix of the electrons. The results of Paper \,I
have been compared with those obtained using the traditional methods
with various electron basic functions.
The latter methods are simpler for calculations, while 
the density matrix formalism enables one to choose
which basis is more adequate at given parameters
of stellar matter. Paper~I has presented also
efficient numerical techniques for calculating the
electron transport coefficients including
the Debye--Waller factor and demonstrated that
this factor is 
much more significant when the field is strongly quantizing. 

In this paper we obtain practical formulae
for a simple and rapid evaluation
of the longitudinal transport
coefficients using the theory developed in Paper~I.

\section{Physical conditions}                             
\label{sect2}
We consider degenerate layers of an outer neutron star
crust (at densities $\rho \la 4 \times 10^{11}$~g~cm$^{-3}$,
below the neutron drip).
Matter of these layers consists of electrons and ions.
We study not too low densities
(see below), at which the electrons are nearly free, and the
ionization is complete due to the high electron pressure.
The ions give the major contribution into the density
while the electrons -- into the pressure.
For simplicity, we mainly consider one component plasma
of ions (=nuclei ($A$, $Z$)); the electron and
ion number densities are related as $n_{\rm e}=Z n_{\rm i}$.
A comprehensive study of thermodynamic properties
of magnetized neutron star crusts
has been performed recently by R\"{o}gnvaldsson et al.\ (1993)
using the Thomas--Fermi approximation. Since we avoid
low densities, we base our consideration on the
approximation of free electrons. This will enable us
to evaluate the transport properties of matter. We will
include the effects of non-ideality of electrons
in a phenomenological manner in Sects.~5 and 6.

Let $\mu$ be the electron chemical potential (including
the electron rest-mass energy, $mc^2$).
The domain of strongly degenerate
electrons we are interested in corresponds to
$T \ll T_{\rm F}$, where
$T_{\rm F}=(\mu - mc^2)/\kB $ is the degeneracy
temperature and \kB\ is the
Boltzmann constant. If $(\mu - mc^2) \ll mc^2$, 
then the electron gas is non-relativistic, while for
$\mu \gg mc^2$ it becomes ultra-relativistic.

In the absence of the magnetic field,
one has the familiar result
\begin{equation}
     n_{\rm e} = {1\over\pi^2 \hbar^3}
      \int_0^\infty f(\ep) p^2 {\rm d} p,
\label{mu_nonmag}
\end{equation}
where
\begin{equation}
    f(\ep) = \left[ \exp \left(
         {\ep - \mu \over k_{\rm B} T} \right)+1 \right]^{-1}
\label{FermiDirac}
\end{equation}
is the Fermi--Dirac distribution function, and
$\ep=c\sqrt{(mc)^2+p^2}$ is the electron energy.
In the case of strong degeneracy, one has
$\mu \approx \mu_0$, where
\begin{equation}
      \mu_0 = \sqrt{m^2c^4+c^2 p_{\rm F0}^2},~~~
      p_{\rm F0} = \hbar (3 \pi^2 n_{\rm e})^{1/3},
\label{mu_0}
\end{equation}
and $p_{\rm F0}$ is the field-free electron Fermi momentum 
  ($p_{\rm F0}/(mc) \approx 1.009 (\rho_6 Z /A)^{1/3}$, 
  $\rho_6$ being density in units
  of 10$^6$ g~cm$^{-3}$). 
The appropriate degeneracy temperature is 
plotted in Fig.~1 by short-dashed line.
For small $\rho$ and $T$, the effects of incomplete ionization
and electron gas non-ideality become important. This domain
is shown schematically by 
the short-dashed line $N$. The line corresponds to
$\bar{\ep} = |\ep_a|$, where $\bar{\ep}$ is
the mean energy per electron in the free electron gas,
and $\ep_a$ is the mean energy
per electron for isolated atoms in the Thomas--Fermi approximation
(e.g., Landau \& Lifshitz 1976).

     \begin{figure}
\begin{center}
\leavevmode
\epsfysize=88mm 
\epsfbox{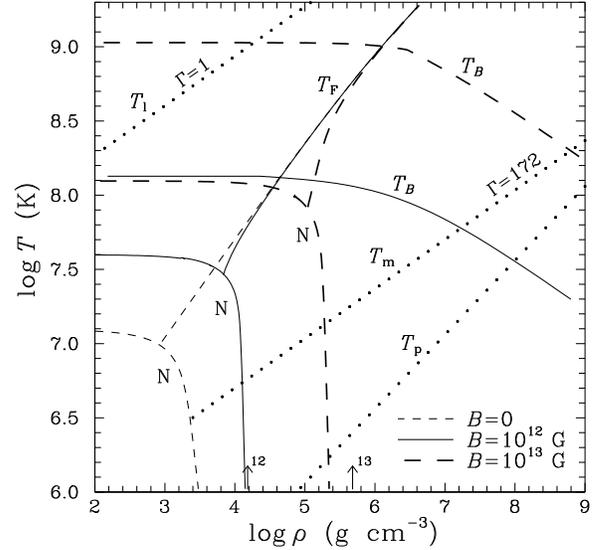}
\end{center}
     \caption[ ]{
   $\rho$ -- $T$ diagram of $^{56}$Fe matter for
        several magnetic fields strengths.
   $T_{\rm F}$ -- electron degeneracy temperature,
   $T_{\rm l}$ -- ion coupling temperature ($\Gamma=1$),
   $T_{\rm m}$ -- melting temperature,
   $T_{\rm p}$ -- ion plasma temperature,
   $T_B$ -- electron quantization temperature;
   arrows indicate strongly quantizing densities $\rho_B$.
   Lines $N$ restrict the low-$T$ low-$\rho$ domain of
   incomplete ionization and electron gas non-ideality.
    $T_{\rm F}$, $T_{\rm B}$, and $N$ curves depend 
    on the magnetic field.
    Dashes show the curves for $B=0$,
    solid lines for $B=10^{12}$~G, and thick long-dash
    lines for $B=10^{13}$~G.
}
     \label{fig1}
     \end{figure}

Let the magnetic field $\vect{B}$ be directed along
the $z$-axis. We shall use the Landau gauge of the vector
potential: $\vect{A}=(-By,0,0)$. Quantum states of a free
electron can be labelled by four
quantum numbers $(p_x,p_z,n,s)$, where $p_x$ determines
the $y$-coordinate of the electron guiding center,
$y_B= p_x/(m \omega_B)$, $p_z$
is the longitudinal electron momentum, $n$=0,1,2,\ldots
enumerates the Landau levels,
and $s$ is a spin variable. Here
$\omega_B = eB/(mc)$ is 
the electron cyclotron frequency, 
and $(-e)$ is the electron charge.
The ground Landau level $n=0$ is non-degenerate with respect
to spin while the  excited levels $n>0$ are
doubly degenerate. Various sets of electron basic wave functions
are analysed, e.g., in Paper~I.
An energy $\ep$\ of a relativistic
electron in the magnetic field is
\begin{eqnarray}
    \ep & = & \ep_n \equiv Emc^2
  \nonumber \\
          & 
        = 
          & 
        \sqrt{m^2 c^4 + c^2 p_z^2 + 2 mc^2 \hbar \omega_B n},
\label{E}
\end{eqnarray}
where $E$ is the natural dimensionless energy.

The number density
of free electrons in a magnetic field is
\begin{eqnarray}
       n_{\rm e} & = & {m \omega_B \over (2 \pi \hbar)^2} \;
             \int_{-\infty}^{+\infty}
             {\rm d} p_z  \, \sum_{n,s} f(\ep_n)
  \nonumber \\
    & 
        =  
          & 
        \int_{mc^2}^{\infty} {\cal N}_{\rm e}(\ep) \,
             \left(- {\partial f \over \partial \ep} \right)
             {\rm d} \ep,
\label{n_e} \\
     {\cal N}_{\rm e}(\ep) & = & {2 m \omega_B \over (2 \pi \hbar)^2}
            \sum_s \sum_{n=0}^{n(E)} p_n(\ep).
\label{N_e}
\end{eqnarray}
Here $p_n(\ep) \equiv mc P_n(E)$ is the largest $z$-momentum of
an electron with energy $\ep$ at a given Landau level,
$n(E)= {\rm Int}(\nu)$ is the
highest Landau level occupied by an electron with energy $\ep$,
\begin{equation}
   P_n(E)= \sqrt{E^2-1-2bn},~~~ \nu=(E^2-1)/(2b),
\label{nu}
\end{equation}
Int$(\nu)$ is integer part of $\nu$, and $b$ is the dimensionless
magnetic field expressed in units of the `critical field' $B_c$
($\hbar \omega_{Bc}=mc^2$),
\begin{equation}
     b={B \over B_c},~~
     B_c = { m^2 c^3 \over e\hbar} = 4.414 \times 10^{13}~{\rm G}.
\label{b}
\end{equation}

Mass density $\rho$
equals $\rho \approx m_{\rm i} n_{\rm i} \approx
(A/Z) m_{\rm u} n_{\rm e}$, where $m_{\rm i}$ is the ion mass,
$m_{\rm u} = 1.660 \times 10^{-24}$~g is the atomic mass unit,
and $A/Z$ is the number of baryons per electron.
Equation~(\ref{n_e}) determines $n_{\rm e}$ as a function
of $T$, $\mu$ and $B$.

Simple analysis of Eq.~({\ref{n_e}) yields the following.
Strongly degenerate electrons populate
the only ground Landau level when density
is not too high, $\rho < \rho_B$, where
$\rho_B = 2.07 \times 10^6 \, b^{3/2} (A/Z)$~g~cm$^{-3}$.
In this case
the Fermi momentum is 
\begin{equation}
     p_{\rm F} = {2 \pi^2 \hbar^2 n_{\rm e} \over m \omega_B}
               = {2 \over 3 b}  p_{\rm F0}
               \left( { p_{\rm F0} \over mc } \right)^2.
\label{p_FB}
\end{equation}
In the domain of $\rho \ll \rho_B$
the magnetic field strongly reduces the electron Fermi
energy and widens the region of incomplete
ionization and electron gas non-ideality (Yakovlev 1984,
Hernquist 1984, Van Riper 1988), 
as shown in Fig.~1 by solid and long-dashed lines. 
In this case, lines $N$ correspond to 
$\bar{\ep} = |\ep_a|$, where $\ep_a$ 
depends on $B$ (R\"{o}gnvaldsson et al.\ 1993).

If $B$ is fixed, then $\mu$ increases with $\rho$,
and degenerate electrons populate higher Landau levels $n$.
If $\rho \gg \rho_B$,
many Landau levels are populated,
and the Fermi energy is almost independent of $B$
($p_{\rm F} \approx p_{\rm F0}$).

It is convenient to introduce
the temperature
\begin{equation}
     T_B = {\hbar \omega_B^\ast \over \kB }
         \approx 1.34 \times 10^8 \, B_{12}
         {mc^2 \over \ep^\ast}~~{\rm K},
\label{T_B}
\end{equation}
where 
$\omega_B^\ast = eBc/\ep^\ast$
is the gyrofrequency of
an electron 
with the typical energy
$\ep^\ast = \max(\bar{\ep},\mu)$,
and $B_{12} = B/(10^{12}~{\rm G})$.
When $T \gg T_B$ the electrons occupy many
Landau levels for any $\rho$ due to high thermal energy.
In this case the thermal widths of the Landau energy levels
($\sim \kB T$) are higher than the inter-level spacing, and
the magnetic field acts as {\it non-quantizing}, regardless
of the electron degeneracy.

In the domain of $T \la T_B$ and $\rho < \rho_B$
(separated into the subdomains of degenerate and non-degenerate
electron gas) the electrons populate mostly the
ground Landau level. In this domain, the magnetic field
is {\it strongly quantizing}, and it affects significantly
all properties of matter (e.g., Van Riper 1988,
Yakovlev \& Kaminker 1994).

Finally, in the domain of $T \la T_B$ and $\rho \gg \rho_B$
the electrons are degenerate and populate many Landau
levels but the inter-level spacing exceeds $\kB T$. Then
the magnetic field is {\it weakly quantizing}.
It does not affect noticeably the bulk properties (pressure,
electron chemical potential) determined by all the
electron Fermi sea but it affects the transport
properties determined by thermal electrons near the
Fermi level.

The state of ions can be characterized by the
ion coupling parameter
\begin{equation}
   \Gamma = {Z^2 e^2 \over a \kB T}
          = 2.275 \, {Z^2 \over T_7}
            \left({ \rho_6  \over A} \right)^{1/3} ,
\label{Gamma}
\end{equation}
where $a =[3/(4 \pi n_{\rm i})]^{1/3}$ is the ion sphere radius,
and $T_7$ is temperature in units of 10$^7$~K.

At high $T$, when $\Gamma \ll 1$, the ions form
a classical Boltzmann gas. With decreasing $T$, the gas
gradually (without any phase transition) becomes a
Coulomb liquid. The liquid is formed
(Hansen 1973) at $\Gamma \approx 1$,
i.e., at $T \approx T_{\rm l}$ (Fig.~1). The liquid
solidifies into the Coulomb crystal (Nagara et al.\ 1987)
at $\Gamma = 172$ ($T=T_{\rm m}$). According to Fig.~1, 
we always have strongly coupled Fe ions, if electrons are 
degenerate ($T_{\rm l} > T_{\rm F}$). However,
for light ions, $T_{\rm l}$ can be lower than $T_{\rm F}$.

At low $T$, zero-point ion vibrations become important
in ion motion. These effects are especially
pronounced if $T \ll T_{\rm p}$, where
\begin{equation}
      T_{\rm p}  =  \hbar \omega_{\rm p} / \kB
          \approx 7.832 \times 10^6 (Z/A)\rho_6^{1/2}~{\rm K},
\label{T_p}
\end{equation}
and
$ \omega_{\rm p}  = \sqrt{ 4 \pi Z^2 e^2 n_{\rm i} / m_{\rm i} } $
is the ion plasma frequency.
The amplitude of zero-point vibrations is commonly much smaller than
the typical inter-ion distance, $a$. With increasing $\rho$,
the amplitude-to-$a$ ratio becomes larger, so that
the vibrations can prevent crystallization
at high $\rho$ (Mochkovitch \& Hansen 1979, Ceperley \& Alder 1980).
This effect is especially pronounced for H and He.
Note that
the Debye temperature of the Coulomb bcc crystal is
$T_{\rm D} =0.45 T_{\rm p}$ (Carr 1961).

We assume that the magnetic fields do not affect the properties
of the ion plasma component, for instance, the phonon spectrum of
the crystal. This is so (e.g., Usov et al.\ 1980) if
$\omega_B = ZeB/(m_{\rm i}c) \ll \omega_{\rm p}$, i.e., if
$B \ll 10^{14} \sqrt{\rho_6}$~G.

\section{Transport properties}                           
\label{sect3}
\subsection{Electron scattering mechanisms}            
The electron transport properties
are determined by
the electron scattering mechanisms.
Consider three important cases
when the electron scattering is
almost elastic (an energy transfer in a collision event
is $ \ll \kB T$):
(i) the Coulomb scattering
on ions in the liquid or gaseous phase ($T > T_{\rm m}$, Fig.~1),
(ii) the scattering on high-temperature phonons
($ T_{\rm D} \la T < T_{\rm m}$);
(iii) the Coulomb scattering
on charged impurities in the lattice
(important for $T \ll T_{\rm D}$).
The impurities represent
ions of charge $Z_{\rm imp} \neq Z$ immersed accidentally in
lattice sites.
Accordingly, our results cover a wide range of temperatures.

The electron-ion
scattering
can approximately be described (e.g., Yakovlev 1984) 
by the Debye-screened Coulomb potential.
Its Fourier image $U(\vect{q})$ is given by
\begin{equation}
   \left| U_{\rm ion}(\vect{q})\right|^2 =
   \left( { 4\pi Ze^2 \over q^2+q_{\rm s}^2} \right)^2,~~~
   (q^{\rm ion}_{\rm s})^2=q_{\rm i}^2+q_{\rm e}^2,
\label{U_ion}
\end{equation}
where $q_{\rm s}$ is an effective screening wavenumber
(inverse screening length),
$q_{\rm i}$ and $q_{\rm e}$
are, respectively, the ion and electron screening wavenumbers.
According to Yakovlev \& Urpin (1980) and
Yakovlev (1984),
\begin{equation}
   q_{\rm i}^{-2} =   
         (2a/3)^2 + r_{\rm D}^2,~~~~
         q_{\rm e}^2= 4 \pi e^2\; ( \partial n_{\rm e}  / \partial \mu),
\label{q_i}
\end{equation}
where 
$r_{\rm D}$ is the
Debye ion radius, $r_{\rm D} = v_{\rm i} / \omega_{\rm p}$,
$v_{\rm i} = \sqrt{ \kB T / m_{\rm i} }$ is the thermal ion velocity.
The Debye radius dominates
($q_{\rm i} \approx 1 /r_{\rm D}$)
in the gaseous regime
$T \gg T_{\rm l}$, while $q_{\rm i} \approx 1.5/a$
in the most important case of strongly coupled
ion liquid ($T_{\rm m} < T \la T_{\rm l}$).
Equation (\ref{q_i}) allows us
to reproduce the $B=0$ transport properties using
the model potential (\ref{U_ion}) instead of
the exact plasma-screened Coulomb potential.

For the scattering on high-temperature phonons,
one has (e.g., Paper\,I)
\begin{equation}
   \left| U_{\rm ph}(\vect{q}) \right|^2 =
   \left( {4\pi Z e^2 \over q} \right)^2\,{r_T^2\over 3}\,
   \exp \left[ -2W(\vect{q}) \right],
\label{U_ph}
\end{equation}
where $r_T^2 = u_{-2} a^2 / \Gamma$
is the mean squared thermal displacement of ions
from their lattice sites,
$u_{-2}$ is a numerical factor determined by the phonon
spectrum ($u_{-2}=13.0$ for the 
body-centered cubic (bcc) lattice,
see, e.g., Mochkovitch \& Hansen 1979),
${\rm e}^{-2W}$ is the Debye--Waller factor,
$ 2W(\vect{q}) \approx (r_T q)^2/3$.
Theoretical formalism for incorporating
this factor into the transport properties
of magnetized matter was developed in Paper\,I.
Note that Eq.~(\ref{U_ph}) is valid for
$T_{\rm D} \la T < T_{\rm m}$.

The scattering on phonons remains to be the dominant electron
scattering mechanism at lower temperatures,
$T \ll T_{\rm D}$.
The low-$T$ phonon scattering is inelastic
(electron energy transfer in a collision is $\sim \kB T$),
and, strictly speaking, it cannot be described using the
formalism of Paper\,I. However this scattering can
approximately (semi-quantitatively) be treated as elastic
(Yakovlev \& Urpin 1980, Raikh \& Yakovlev 1982) as long as
the Umklapp processes dominate over
the normal processes of electron-phonon interaction.
The Umklapp processes are known to be most important
at high temperatures, under typical conditions
in neutron star crusts. However they are
``frozen out'', and the electron scattering becomes
essentially inelastic.
For instance, for nonquantizing fields, this happens
at $T \ll T_U \sim T_{\rm p}
Z^{1/3} e^2/(\hbar v_{\rm F})
\approx  [Z^{2/3}+0.98(A/\rho_6)^{2/3}]^{1/2} (T_p/137)
$,
where $v_{\rm F}$ is the electron Fermi velocity.
We propose to extend the formalism of high-temperature
phonons to
$T  \ll T_{\rm D}$ and obtain, thus,
the reliable estimates of low-temperature transport properties
at $T_U \la T \ll T_{\rm D}$.
For this purpose, as can be shown, for instance, from
the results of Raikh \& Yakovlev (1982) and Baiko \& Yakovlev
(1995), the scattering potential
(\ref{U_ph}) should be modified in two ways.
First, the pre-exponent factor $r_T^2$ should be
replaced by $r_{T1}^2 = r_T^2 
G(t)
$, where
$G(t)=t/\sqrt{t^2+t_0^2}$ describes the reduction
of thermal ion displacements at low $T$, 
$t \equiv T/T_{\rm p}$,
and $t_0\sim 0.1$ is a numerical parameter ($t_0=0.132$ 
for the bcc lattice). 
Second, one should remind that the Debye--Waller factor
is determined by the total (thermal + zero point)
mean squared ion displacement $r_{T2}^2$ which reduces to the
thermal squared displacement $r_T^2$ at high $T$. Thus we
should replace $r_T^2$
by $r_{T2}^2 = r_T^2 \{1+ \exp(-9.1t)[u_{-1}/(2 u_{2} t)] \}$
in the Debye--Waller exponent (Baiko \& Yakovlev 1995) for it to be
accurate at low $T$. In this case, $u_{-1}$ is another parameter
of the phonon spectrum ($u_{-1} = 2.800$, for the bcc lattice).
The above modifications do not
violate the formalism of Paper\,I.

Finally, the scattering on impurities corresponds to
(e.g., Yakovlev \& Urpin 1980)
\begin{equation}
   \left| U_{\rm imp}(\vect{q})\right|^2 =
   \left[{ 4\pi (Z_{\rm imp}-Z)e^2 \over q^2+q_{\rm s}^2 } \right]^2.
\label{U_imp}
\end{equation}
In this case the screening wavenumber is
$(q^{\rm imp}_{\rm s})^2=q_{\rm e}^2+q_{\rm imp}^2$, where
$q_{\rm e}$ is given by Eq.\,(\ref{q_i}), and $q_{\rm imp}$
is an inverse impurity correlation length.
The scattering on impurities is similar
to that on ions. This scattering acts
at $T<T_{\rm m}$, just as the scattering on phonons, but
actually it dominates at very low $T$ (see Yakovlev \& Urpin 1980). 

\subsection{Transport coefficients}                  
Let ${\cal E}$,  $\partial \mu / \partial z$ and
$\partial T / \partial z$ be, respectively, weak
and locally constant
electric field, electron chemical potential gradient,
and temperature gradient along
$\vect{B}$. They induce the electron electric and thermal
currents with the current densities ${\cal J}$ and ${\cal Q}$:
\begin{eqnarray}
    {\cal J} & = & \sigma \left( {\cal E} +
    {1 \over e} {\partial \mu \over \partial z} \right)
    + \beta {\partial T \over \partial z},
  \nonumber \\
    {\cal Q} 
       & 
    = 
         & 
    - \beta T \left( {\cal E} + {1 \over e}
    {\partial \mu \over \partial z} \right)
    - \lambda {\partial T \over \partial z}.
\label{Transport1}
\end{eqnarray}
Here $\sigma$ is the longitudinal electric conductivity,
while $\beta$ and $\lambda$ are two other auxiliary
transport coefficients. The appearance
of $\beta$ in the expressions for ${\cal J}$ and ${\cal Q}$
reflects the Onsager symmetry principle. For practical use,
the expressions (\ref{Transport1}) can be rewritten as
\begin{equation}
   {\cal E} + {1 \over e} {\partial \mu \over \partial z}
    = {{\cal J} \over \sigma} - \alpha
    {\partial T \over \partial z},
    ~~~~~
    {\cal Q} = - \alpha T {\cal J} -
    \kappa {\partial T \over \partial z},
\label{Transport2}
\end{equation}
where
\begin{equation}
   \alpha = \beta / \sigma,
   ~~~~\kappa = \lambda - T (\beta^2 / \sigma)
\label{kappa}
\end{equation}
are, respectively, the longitudinal
thermopower, and thermal conductivity.
The transport coefficients $\sigma$, $\beta$, and $\lambda$
are convenient for calculation
(see Eq.~(\ref{Int_Phi}) below) while
$\sigma$, $\alpha$ and $\kappa$ directly enter
the equations which govern the distributions of temperature
and magnetic field in neutron stars
(e.g., Urpin \& Yakovlev 1980).
The coefficients $\sigma$, $\kappa$, and
$\alpha$ determine fully the electron transport of charge and heat
along $\vect{B}$.

Let $\rho_{ns's}(z,p_z)$ be the spin polarization
density matrix (Paper\,I) of electrons for the case
when the electron gas is slightly non-uniform along $\vect{B}$.
Here $s$ and $s'$ are the spin variables (Sect.~2).
In the linear regime, deviations from the equilibrium
are small. The zero-order
density matrix is
$f(\ep) \delta_{ss'}$, where $f(\ep)$
is a local Fermi--Dirac distribution (\ref{FermiDirac})
which depends on $z$ parametrically through $\mu$ and $T$.
In the first approximation, according to Paper\,I,
\begin{eqnarray}
      \rho_{n ss'}(z,p_z) & = &
      f(\ep) \delta_{ss'} +
      \eta \,l\,
      {\partial f(\ep)\over \partial \ep}\,
      \left[e{\cal E}+{\partial\mu\over\partial z}+
           \right.
     \nonumber \\
            && \left.
      {\ep - \mu \over T}\,{\partial T\over\partial z}
      \right]\,\varphi_{\eta n s s'}(\ep),
\label{DensityMatrix}
\end{eqnarray}
where $\eta=$sign$(p_z)$, and the functions
$\varphi_{\eta n s s'}(\ep)$ determine
non-equilibrium corrections to the density matrix;
$l$ is an electron scattering
length:
\begin{eqnarray}
    l_{\rm ion} & = & {m c^2 \hbar \omega_B \over 2 \pi
    n_{\rm i} Z^2 e^4},
    ~~~~~~
    l_{\rm ph} = {3 \over 4 \pi n_{\rm i}}
    \left({\hbar c \over Z e^2 r_{T1}} \right)^2,
     \nonumber \\
    l_{\rm imp} 
         & 
   = 
       & 
   {m c^2 \hbar \omega_B \over 2 \pi
    n_{\rm imp} (Z_{\rm imp}-Z)^2 e^4},
\label{l}
\end{eqnarray}
$n_{\rm imp}$ being the number density of impurities.
The set of equations for $\varphi_{\eta n s s'}(\ep)$
has been derived in Paper\,I.
Paper\,I presents also the mathematical formalism
for solving these equations.
After introducing the scale lengths
(\ref{l}), the equations describing the scattering on ions and
impurities appear to be formally the same, i.e., there
is no need to consider these scatterings separately.
We shall refer to the scattering on ions and impurities
as the {\it Coulomb} (C) scattering.
If the functions $\varphi_{\eta n s s'}(\ep)$ are
found, the transport coefficients in Eqs.\,(\ref{Transport1})
can be calculated as (Paper\,I)
\begin{eqnarray}
   \left(
   \begin{array}{c}
       \sigma \\ \beta \\ \lambda
   \end{array}
   \right)
   & = & {2 m \omega_B l \over (2 \pi \hbar)^2 }
   \int_{m c^2}^\infty\!
   \left(
   \begin{array}{c}
       e^2 \\ e(\ep-\mu)/T \\ (\ep -\mu)^2/T
   \end{array}
   \right)
    \nonumber \\
       & 
       \times 
       &
   \Phi(\ep)
   \left( -{\partial f \over \partial \ep} \right)
   {\rm d} \ep,
\label{Int_Phi}
\end{eqnarray}
where
\begin{equation}
     \Phi(\ep) = {1 \over 2} \sum_{\eta=\pm 1}
     \sum_{n=0}^{n(E)} \sum_{s=\pm1} \varphi_{\eta nss}(\ep).
\label{Phi}
\end{equation}

It is also convenient to introduce the function $\Psi(E)$,
\begin{equation}
    \Psi_{\rm C}(E)= b^2 \Phi_{\rm C}(\ep),~~~
    \Psi_{\rm ph}(E)= b \Phi_{\rm ph}(\ep).
\label{Psi}
\end{equation}

Equations (\ref{Int_Phi}) can be written in a more transparent
form if we introduce the effective energy dependent
electron relaxation time:
\begin{equation}
    \tau(\ep)  =  { \ep l m \omega_B \over 2 (\pi \hbar c)^2
         {\cal N}_{\rm e}(\ep) } \Phi(\ep),
\label{tau}
\end{equation}
where ${\cal N}_{\rm e}$ is given by Eq.~(\ref{N_e}).
Then
\begin{eqnarray}
   \left(
   \begin{array}{c}
       \sigma \\ \beta \\ \lambda
   \end{array}
   \right)
   & = &
   \int_{m c^2}^\infty\!
   \left(
   \begin{array}{c}
       e^2 \\ e(\ep-\mu)/T \\ (\ep -\mu)^2/T
   \end{array}
   \right)
    \nonumber  \\
        &  \times &
   { {\cal N}_{\rm e}(\ep) \tau(\ep)c^2 \over \ep}
   \left( -{\partial f \over \partial \ep} \right)
   {\rm d} \ep ,
\label{Int_tau}
\end{eqnarray}
where the energy integration represents
the statistical averaging of the relaxation time
with the energy derivative of the Fermi--Dirac
distribution. Note that Eqs.~(\ref{Int_Phi})
and (\ref{Int_tau}) are valid for any electron
degeneracy.
Note also that Eqs.~(19) and (41) of Paper\,I should contain
${\cal N}_{\rm e}(\ep)$ instead of $n_{\rm e}$.

Generally, $\tau(\ep)$ is an oscillating function of
electron energy as discussed in Sect.~4.
Let us mention one important case when
$\tau(\ep)$ and
${\cal N}_{\rm e}(\ep)$ vary with $\ep$
much slower than $\partial f(\ep) / \partial \ep$.
Then, for strongly degenerate electrons,
one obtains the results which look formally similar to those for $B=0$:
\begin{eqnarray}
     \sigma  & \approx & {e^2 c^2 n_{\rm e} \tau(\mu) \over \mu},~~
     \kappa   \approx  \lambda \approx
     {\pi^2 \kB^2 T  \over 3 e^2}\, \sigma,
     \nonumber \\
     \alpha 
          & 
   \approx  
        & 
   {\pi^2 \kB^2 T \over 3 e} \;
         { \partial \over \partial \ep} \;
         \left. \ln \left( {\cal N}_{\rm e}(\ep)
         \tau(\ep) \over \ep \right)
         \right|_{\ep=\mu},
\label{Transport0}
\end{eqnarray}
with $n_{\rm e}={\cal N}_{\rm e}(\mu)$.

The equations of this section are equally valid for the quantizing and
non-quantizing magnetic fields. 

\section{Relaxation time $\tau(\ep)$ (or $\Psi(E)$)}        

\subsection{Quantum oscillations}                            
Evaluation of the transport coefficients $\sigma$, $\alpha$, and
$\kappa$ consists of two stages. First, the equations of Paper\,I for
$\varphi_{\eta n s s'}(\ep)$ should
be solved and the function $\Phi(\ep)$
(or, equivalently, $\Psi(E)$, or $\tau(\ep)$) determined.
Second, the energy integrations (\ref{Int_Phi}) have to be performed,
and the transport coefficients (\ref{kappa})
found. The second stage corresponds actually to
the energy averaging of the relaxation time;
it will be analysed in Sect.~5.
Here we consider the first stage.

The main problem is to reproduce correctly {\it quantum
oscillations} of $\Psi(E)$ or $\tau(\ep)$ which
occur since 
electrons
of energy $\ep$ 
populate
new Landau levels with growing $\ep$. Population
of an $n$-th level takes place when the energy variable $\nu$
given by Eq.~(\ref{nu}) 
exceeds $n$.
The oscillations originate from the
square root singularities of the density of states
of free electrons in a magnetic field. Just
behind the threshold for a given level ($0 <(\nu - n) \ll 1$)
$\Phi(\ep)$ and $\tau(\ep)$ behave
as $\sqrt{\nu-n}$ which makes the oscillations
important especially for not too high $n$.

Below we obtain practical equations for $\Psi(E)$
or $\tau(\ep)$ at various electron energies.

\subsection{Semiclassical approach for $\nu \gg 1$}      
If $\nu \gg 1$ (electrons
occupy many Landau levels),
$\Psi(E)$ can be calculated numerically, using the
technique of Paper\,I, but it cannot be
expressed in a closed analytic
form. We will derive
(Sect.~4.4) accurate fitting expressions
based on the
semiclassical approach described below.

When many Landau levels are occupied,
the relaxation time
$\tau(\ep)$ is expected to be close to the familiar
classical non-magnetic quantity
\begin{equation}
   \tau^{-1}(\ep) = n_{\rm i} v_0 \sigma_{\rm tr}(\ep),
\label{tau_0}
\end{equation}
where $v_0$ is the electron velocity without
magnetic quantization effects ($v_0=p_0 c^2/\ep$,
$p_0=mcP_0=mc \sqrt{E^2- 1}$),
$n_{\rm i}$ is the number density of scatterers,
and $\sigma_{\rm tr}(\ep)$ is the transport
scattering cross section:
\begin{equation}
   \sigma_{\rm tr}(\ep) =
   \int {{\rm d}\Omega\over 4\pi}\int {\rm d}\Omega'
   \sigma(\Theta)\,(1-\cos\Theta).
\label{sigma_tr}
\end{equation}
In this case $\sigma(\Theta)$ is the differential
scattering cross section, $\Theta$ is
the scattering angle,
d$\Omega$ and d$\Omega'$ are
the solid angle elements of non-quantized electron
momenta $\vect{p}$ and $\vect{p}'$
before and after scattering, respectively.
In the Born approximation,
\begin{equation}
   \sigma(\Theta) =
      {|U(\vect{q})|^2 \ep^2 \over 4\pi^2 \hbar^4c^4}\,
      \left(1-{v_0^2\over c^2}\,\sin^2{\Theta\over 2}\right).
\label{Born}
\end{equation}
Assuming
the isotropic distribution over momentum transfers
$\hbar \vect{q} = \vect{p}'-\vect{p}$
in the non-magnetic case,
Eq.\,(\ref{sigma_tr}) can be rewritten as
\begin{equation}
   \sigma_{\rm tr}(\ep) =
   \int {{\rm d}\Omega\,{\rm d}\Omega'\over 4\pi}\,
   \sigma(\Theta) \, {3\hbar^2q_z^2 \over 2p_0^2}.
\label{sigma_tr1}
\end{equation}

Now consider
weakly quantizing magnetic fields.
Using Eq.\,(\ref{E}), we can introduce
the quantized transverse momentum
$ p_\perp(n) = p_0 \, \sin \vartheta =
\sqrt{2m \hbar \omega_B n}$,
where $n$ is a Landau number, and 
$\vartheta$ is an electron pitch angle.
Accordingly,
every $n$ corresponds to two values of $\vartheta$,
below and above $\pi / 2$. Then an integration
over $\vartheta$ from 0 to $\pi$
can be replaced as:
\begin{equation}
   \int_0^\pi {\rm d}\vartheta \,\sin \vartheta \, \ldots
   \rightarrow
   {m \hbar\omega_B \over p_0 |p_z|}
   \sum_{n=0}^{n(E)} \sum_{\eta=\pm1} \, \ldots
\label{sum_perp}
\end{equation}
At this stage we replace the integrals over $\vartheta$ and
$\vartheta'$ in Eq.\,(\ref{sigma_tr1})
by the sums over $n$, $\eta$
and $n'$, $\eta'$. The dips of $\tau(\ep)$
or $\Phi(\ep)$ behind every Landau threshold
are caused by the density-of-state 
singularity when either $n$ or $n'$ equals
$n(E)$. Since $n(E)$ is assumed to be large,
these terms correspond to $\vartheta \approx \pi/2$ or
$\vartheta' \approx \pi/2$, respectively. The sum over
$\eta$ or $\eta'$ in these terms is equivalent to
introducing a factor 2.
Thus, the term $n=n(E)$ can be evaluated
by setting $\vartheta=\pi/2$ in
Eq.\,(\ref{sigma_tr1}), multiplying by
$2m \hbar \omega_B / (p_0|p_z|)$ and integrating over all remaining
variables. The term $n'=n(E)$ is similar. Then
\begin{equation}
    \sigma_{\rm tr}(\ep) =
    \sigma_{\rm cl}(\ep)\, +
    {3m \hbar \omega_B \over p_0 |p_z| }\,\sigma_{\rm q}(\ep),
\label{sigma_tr2}
\end{equation}
where
\begin{eqnarray}
    \sigma_{\rm cl}(\ep) &=&
    \int {\rm d} \Omega' \,  \sigma(\Theta) \,
   (1- \cos \Theta),
\label{sigma_cl} \\
    \sigma_{\rm q}(\ep) &=&
    \int {\rm d} \Omega' \,  \sigma(\Theta) \,
    \cos^2\vartheta' =
    \int {\rm d} \Omega' \sigma(\Theta) \,
    { \sin^2 \Theta \over 2}.
\label{sigma_q}
\end{eqnarray}
The last term in
Eq.\,(\ref{sigma_tr2}) comes from the terms $n=n(E)$,
$n'=n(E)$. Its denominator contains
the longitudinal momentum $p_z$ at $n=n(E)$ which
vanishes just at a new Landau threshold
and produces the required quantum oscillations.
The first term, $\sigma_{\rm cl}(\ep)$,
is non-oscillating. Actually it should
be somewhat lower since
we have subtracted
the oscillating term 
from $\sigma_{\rm tr}$ but we neglect
this difference in the present section. Equation~(\ref{sigma_tr2})
can also be inaccurate for small momenta transfers associated
with the transitions between discrete neighboring $n$ and $n'$
(which require a more detailed consideration). In order
to allow for this effect, we introduce the lower
momentum transfer cutoff $q=q_{\rm min}$ while calculating
$\sigma_{\rm cl}(\ep)$ and $\sigma_{\rm q}(\ep)$.
We specify $q_{\rm min}$ in Sect.~4.4.

From Eq.~(\ref{U_ion}), for the scattering on
ions we obtain
\begin{equation}
   \sigma_{\rm cl,q}^{\rm ion}(\ep) =
   4\pi\left({Ze^2\over p_0 v_0}\right)^2
   R_{\rm cl,q}^{\rm C}(E,y),
\label{sigma_ion}
\end{equation}
where
\begin{eqnarray}
   R_{\rm cl}^{\rm C}(E,y)
     & = & \Lambda_1
     - {v_0^2 \over c^2} \Lambda_2,
\nonumber \\
   R_{\rm q}^{\rm C}(E,y)
     & = & R_{\rm cl}^{\rm C}(E,y)
         - \Lambda_2 +
         {v_0^2 \over c^2} \Lambda_3,
\nonumber \\
   \Lambda_{k}
     & \equiv & {1 \over 2}
     \int_y^1 \,{z^k \, {\rm d}z \over (z+u)^2},
\nonumber \\
    \Lambda_1 & = & {1 \over 2}
     \ln \left( {1+u \over y+u} \right) - {u (1-y) \over 2(1+u)(y+u)},
\nonumber \\
    \Lambda_2 & = & {1-y \over 2}
     -
    u \ln \left( {1+u \over y+u} \right) + {u^2 (1-y) \over 2(1+u)(y+u)},
\nonumber \\
    \Lambda_3 & = & {1-y^2 \over 4} - u (1-y)
     \nonumber \\
        & 
   + 
        & 
   {3 \over 2} u^2 \ln \left( {1+u \over y+u} \right)
    - {u^3 (1-y) \over 2 (1+u)(y+u)},
\label{R_Coul}
\end{eqnarray}
$u=[\hbar q_{\rm s}/(2 p_0)]^2$ is the screening parameter,
and $y=[\hbar q_{\rm min}/(2 p_0)]^2$.

For the scattering on 
impurities,
$\sigma_{\rm cl}^{\rm imp}(\ep)$ and
$\sigma_{\rm q}^{\rm imp}(\ep)$ are obtained
from Eq.\,(\ref{sigma_ion}) by replacing
$Z \to (Z_{\rm imp}-Z)$.

In the case of the scattering on phonons,
from Eqs.~(\ref{U_ph}) and (\ref{sigma_tr2}) we obtain 
\begin{equation}
   \sigma_{\rm cl,q}^{\rm ph}(\ep,y)  =
   {8\pi\over 3} \left({Ze^2\over\hbar v_0}\right)^2
   r_{T1}^2 \,R_{\rm cl,q}^{\rm ph}(E,y),
\label{sigma_ph}
\end{equation}
where
\begin{eqnarray}
   R_{\rm cl}^{\rm ph}(E,y) & = & L_1- L_2 \,{v_0^2\over 2c^2},
\nonumber \\
   R_{\rm q}^{\rm ph}(E,y) & = &  R_{\rm cl}^{\rm ph}(E,y) -
   {1 \over 2} L_2
   +  {v_0^2\over 3c^2} \, L_3,
\nonumber \\
   L_k & = & k \int_y^1 z^{k-1} {\rm e}^{-wy} {\rm d}y,
\nonumber \\
   L_1 & = &
    {1 \over w} \left( {\rm e}^{-wy} - {\rm e}^{-w} \right),
\nonumber \\
   L_2 & = &
    {2 \over w^2} \left[(1+wy){\rm e}^{-wy} -
    (1+w){\rm e}^{-w} \right],
\nonumber \\
   L_3 
   & = &
    {3 \over w^3} \left[(2+2wy+w^2y^2){\rm e}^{-wy} 
         \right.
     \nonumber \\
         & 
    - 
         &   
         \left.
    (2+2w+w^2){\rm e}^{-w} \right],
\label{R_ph}
\end{eqnarray}
$w=(4/3)(p_0 r_{T2}/ \hbar)^2$. As explained in Sect.~3.1,
these equations
are strictly valid at
$T_{\rm D} \la T < T_{\rm m}$, and
produce reasonable estimates at
$T_U \la T \ll T_{\rm D}$.

Now, using Eqs.~(\ref{tau_0}) and
(\ref{sigma_tr2}), we can easily express the inverse energy
dependent relaxation time as a sum of the main (non-oscillating) and
oscillating terms.

\subsection{Non-magnetic case}                        
If $B=0$ the relaxation time
of an electron is readily obtained from
Eqs.~(\ref{tau_0}), (\ref{sigma_tr2})
by omitting the oscillating term
and setting $q_{\rm min}=0$ ($y=0$).
Then we immediately recover Eqs.~(23) -- (28)
of Paper\,I for $\tau(\ep)$
due to the Coulomb and phonon scatterings.

The function $\Psi(E)$ is defined by Eq.\,(\ref{Psi}). 
The function
${\cal N}_{\rm e}(\ep)$ which enters Eq.\,(\ref{tau})
and relates $\tau(\ep)$ and $\Psi(E)$
is given by Eq.\,(\ref{N_e}). If $B \to 0$,
${\cal N}_{\rm e}(\ep)$ is easily calculated
by replacing the sum over $n$
with the integral, which yields the trivial result
${\cal N}_{\rm e}(\ep)= (p_0/ \hbar)^3/(3 \pi^2)$.
The appropriate functions $\Psi(E)$ are
\begin{equation}
   \Psi_{\rm C}(E) =
   {P_0^6 \over 3 \Lambda(E) E^2},~~~
   \Psi_{\rm ph}(E)  =   { P_0^4 \over 3 E^2 
       R_{\rm cl}^{\rm ph}(E,0)},
\label{tau_ion0}
\end{equation}
where $\Lambda(E)  =  R_{\rm cl}^{\rm C}(E,0)$
is the Coulomb logarithm given by Eq.~(24) of Paper\,I.

Once the energy dependent relaxation time
$\tau_0(\ep)$ is known, the transport properties
are easily evaluated from
Eq.\,(\ref{Transport0}). In this way we
reproduce familiar transport coefficients
of degenerate electrons for the scattering
mechanisms of study.

\subsection{Fitting expressions for  $\nu >1$}          
We have performed extensive calculations of $\Psi(E)$
for the Coulomb and phonon scattering
potentials, using the formalism of Paper\,I. 
We have mainly used the distribution
function framework with the fixed-spin basis.
According to Paper\,I, this yields sufficient accuracy in most
cases of interest. However, the accuracy has been
additionally controlled using the density matrix formalism.
We have considered a wide range of magnetic fields
$10^{10}~{\rm G} \la B  \la 10^{14}$~G 
sufficient for applications.
In the case of the Coulomb potential, $\Psi_{\rm C}(E)$ depends
also on the screening wavenumber $q_{\rm s}$ (Sect.~3.1).
Instead of $q_{\rm s}$, it is convenient to introduce
the dimensionless screening parameter
$u= [\hbar q_{\rm s}/(2p_0)]^2$, defined in Eq.~(\ref{R_Coul}).
We have calculated $\Psi_{\rm C}(E)$ treating $u$
as a free parameter varied 
from 0 to 1.
These results are equally valid for
the scattering on ions and impurities 
(Sect.~3).
In the case of the scattering on phonons, $\Psi_{\rm ph}(E)$
depends on the Debye--Waller factor defined in Eq.\,(\ref{U_ph}).
The effect of this factor is described by the dimensionless
parameter $w=(4/3)(p_0 r_{T2}/ \hbar)^2$ in accordance
with Eq.\,(\ref{R_ph}).
While calculating $\Psi_{\rm ph}(E)$,
we treat $w$ as a free parameter varied
in a wide interval,
$0 \leq w \la 10$.

Our computations
cover an extended range of electron energies.
The energy variable $\nu$
has been varied up to 100, 
allowing
population of up to 100 Landau energy levels.

We have produced vast tables of $\Psi(E)$ 
(for different $E$, $b$, $u$, $w$). They are
inconvenient for practical use but we
have been able to fit all the results
by analytic formulae. We have started with simple
semiclassical Eqs.~(\ref{tau_0}) and (\ref{sigma_tr2}). They
reproduce the main features of $\Psi(E)$
at large $\nu$ but they appear to be inexact
near the Landau thresholds.
We have managed to modify
Eqs.~(\ref{tau_0}) and (\ref{sigma_tr2}) in such a way
that the resulting formulae fit $\Psi(E)$ accurately
for any parameters $b$, $u$, $w$ and
for $\nu>1$. We have treated the minimum momentum cutoff
$q_{\rm min}$, or $y$ (Sect.~4.2), as a fit parameter.
The fit formula reads ($\nu >1$)
\begin{eqnarray}
    \Psi(E)& = & \sqrt{\Psi_a^2(E)+ \Psi_b^2(E)},
\label{Fit} \\
   \Psi_a(E) & =  & \Psi_0(E) 
   \left[ R_{\rm cl}(E,y) \rule{0mm}{3ex} \right.
     \nonumber \\
         & 
   + 
        & 
   { 3 \over P_0} 
          \left( { b \over P_n} - b^{1/4} \sqrt{2 P_n}  \right)
          \left. \rule{0mm}{3ex} R_{\rm q}(E,y) \right]^{-1},
\label{phi_a} \\
    \Psi_0^{\rm C}(E) & = & {P_0^6 \over 3 E^2},~~~
    \Psi_0^{\rm ph}(E) = {P_0^4 \over 3 E^2},
\label{Psi_0}
\end{eqnarray}
where $y=0.5(P_n/P_0)^2$, $P_n = \sqrt{E^2-1-2bn}$;
$n = n(E) = {\rm Int}(\nu)$ in accordance with Eq.\,(\ref{nu});
the functions $R_{\rm cl}(E,y)$, $R_{\rm q}(E,y)$
are defined in Eqs.\,(\ref{R_Coul}) and (\ref{R_ph}).

The expression for $\Psi_b(E)$
depends on $n$. When $n \geq 3$,
for all scattering mechanisms of study we have
\begin{equation}
    \Psi_b(E)  = 
          { P_0 P_n \Psi_0(E) / (3b) 
            \over
         [ R_{\rm q}(E,y_+) \! + \! R_{\rm q}(E,y_-) ]
          ( 1 \! + \! \nu (P_n / \sqrt{b})^{5/2}) },
\label{phi_b}
\end{equation}
where $y_\pm =0.5[(P_{n-1} \pm P_n)/P_0]^2$. When
$n = 1$ or 2, we obtain
\begin{eqnarray}
    \Psi_b(E)
    & = &  { P_0 P_n \over
          Q_2(\xi_+ + \xi_{\rm s},0,1)+
           Q_2(\xi_- + \xi_{\rm s},0,1)}
     \nonumber \\
         & 
   \times 
      & 
   { b^k \over [(E+1)/2]^2 + [b/(E+1)]^2 },
\label{phi_b1}
\end{eqnarray}
where  $k_{\rm C}=2$ and $k_{\rm ph}=1$ according to Eq.~(\ref{Psi}),
$\xi_\pm = (P_0 \pm P_n)^2/(2b)$,
$\xi_{\rm s}^{\rm C} = (a_{\rm m} q_{\rm s})^2/2$,
$\xi_{\rm s}^{\rm ph}= 0$,
and 
$a_{\rm m}= \sqrt{\hbar c/(eB)}$ 
is the magnetic quantum length.
The functions $Q_2(\xi, n',n)$ have
been introduced in Paper\,I.
We have
\begin{eqnarray}
    Q_2^{\rm C}(\xi,0,1) & = &  \int_0^\infty
       {\zeta \, {\rm e}^{-\zeta} \over (\zeta + \xi)^2 }
        \, {\rm d}\zeta
        =  (1 + \xi) {\rm e}^\xi E_1(\xi) -1,
\nonumber \\
   Q_2^{\rm ph}(\xi,0,1) & = &  \int_0^\infty
       {\zeta \, {\rm e}^{-\zeta g-g\xi+\xi} \over \zeta + \xi }
        \, {\rm d}\zeta
     \nonumber \\
            & 
       = 
            & 
       { {\rm e}^{- g \xi + \xi} \over g }
         \, \left[1 - g \xi {\rm e}^{g \xi}
         E_1(g \xi) \right],
\label{Q_2}
\end{eqnarray}
where $g= 1+w/(4 \nu)$,
and $E_1(\xi)$ is the integral exponent
which is easily calculated (Abramowitz \& Stegun 1972).

Equations (\ref{Fit}) -- (\ref{phi_a}) are valid for the
Coulomb and phonon scatterings. The function $\Psi_a(E)$
is associated with the modified semiclassical expression
of $\Psi(E)$ (Sect.~4.2), whereas $\Psi_b(E)$ is
introduced to improve the fit accuracy just behind a new
Landau level threshold, at $(\nu-n) \ll 1$.
In the limit of $B \to 0$ the equations reproduce
the correct non-magnetic
expression $\Psi(E)=\Psi_0(E)/R_{\rm cl}(E,0)$.

The fitting equations describe
the tables of
$\Psi(E)$ for all values of $b$, $u$, $w$ (see above)
at any energy $E$. They are based on the
correct semiclassical expressions (Sect.~4.2), and their accuracy
does not become worse with
increasing $E$ as demonstrated in Figs.~\ref{fig4.1}
and \ref{fig4.2}. The figures present $\Phi(E)$ versus
$\nu$ for different scattering mechanisms,
magnetic fields $b$, and charge numbers $Z$.
For illustration,
the screening wavenumber $q_{\rm s}$
in the case of the Coulomb scattering
has been calculated assuming strong degeneracy, with 
the Fermi level being equal to the current value of
electron energy, $\mu= \ep$. This allows us to
find the electron number density from Eq.~(\ref{n_e}), and determine
then $q_{\rm s}$ from Eqs.~(\ref{U_ion}) and (\ref{q_i}).
For the scattering
on phonons, we have considered several values of $T$ which enter the
Debye--Waller factor defined in Eq.~(\ref{U_ph}). If
$T_7 \ll 1$, the
Debye--Waller factor is negligible, i.e., ${\rm e}^{-2W}=1$ in
Eq.~(\ref{U_ph}).
Figure~\ref{fig4.1} demonstrates population of the
low-lying Landau levels $n$=1, 2, 3 and 4. Figure~\ref{fig4.2}
displays population of higher levels, $n=20$ and
$n=99$.
The overall accuracy of the fits is around 3\%
which seems to be satisfactory for fitting oscillating
functions. The accuracy becomes worse (up to 8\%
for $n >5$) at $0< \nu - n \ll 1$,
where $\Psi(E)$ contains dips, but 
this is not important: the dips are
easily smeared out due to subsequent energy
averaging of $\Psi(E)$ (Sect.~5).

%
     \begin{figure}[t]
\begin{center}
\leavevmode
\epsfysize=124mm 
\epsfbox{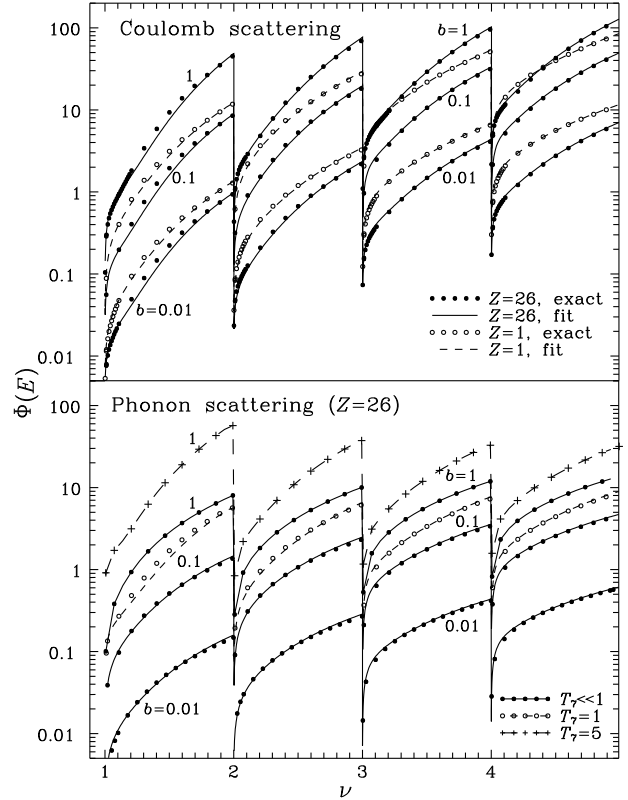}
\end{center}
     \caption[ ]{
   Calculated and fitted function $\Phi(E)$
   versus $\nu$ for Coulomb and phonon
   scatterings at different parameters of matter
   in the case when the electrons populate the
   Landau levels $n$=1,2,3,4.
}
     \label{fig4.1}
     \end{figure}
%
%
\begin{figure}[t]
\begin{center}
\leavevmode
\epsfysize=124mm 
\epsfbox{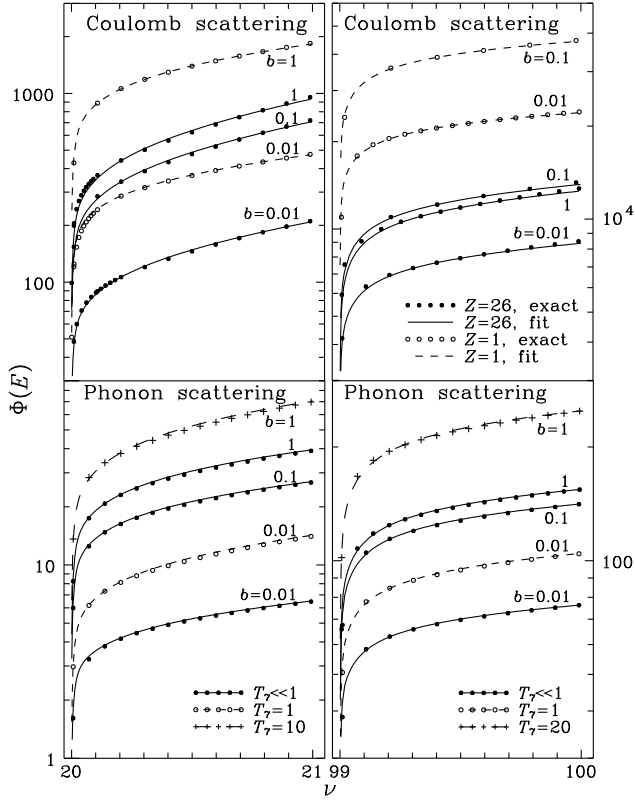}
\end{center}
\caption[ ]{
   Same as in Fig.\ \ref{fig4.1} for population of the
   Landau levels $n$=20 and 99.
}
     \label{fig4.2}
     \end{figure}

Previously the function $\Phi(E)$ was calculated and
fitted, accurately and thoroughly, by Hernquist (1984).
The author considered the Coulomb and high-$T$ phonon
scatterings in iron matter at $B=10^{10}$, $3 \times 10^{10}$,
$10^{11}$, $3 \times 10^{11}$,
$10^{12}$, $3 \times 10^{12}$,
$10^{13}$, $3 \times 10^{13}$,
$10^{14}$~G and included population
of $n<30$ Landau levels.
His fit formula (his Eq. (178))
contained 3 adjustable parameters for every Landau level
(90 fit parameters for every $B$ and each scattering mechanism).
However, Hernquist (1994)
did not take into account the Debye--Waller
factor which made his results for the phonon scattering
not very accurate (Paper\,I). Our fitting
formulae require no tables of fit parameters; they are
valid for any magnetic field and Landau level number
of practical interest, as well as for any chemical element.

\subsection{Low energies, $\nu <1$}                     
If $\nu<1$, the electrons with energy $\ep$ occupy
the ground Landau level $n$=0.
In this case $\Psi(E)$ is expressed
in a closed form using the results of Paper\,I.
For the Coulomb and phonon scatterings, we obtain
\begin{equation}
   \Phi(E)  =  {P_0^2 \over 2Q_2( \xi_+ +\xi_{\rm s},0,0)},
\label{Phi_ground}
\end{equation}
where
$\xi_+ = 2 P_0^2/b$, and $\xi_{\rm s}$ is defined in (\ref{phi_b1}).
The function $Q_2(\xi,0,0)$ is (cf. Eq.~(\ref{Q_2}))
\begin{eqnarray}
   Q_2^{\rm C}(\xi,0,0) & = &  \int_0^\infty
   { {\rm e}^{-\zeta} \over (\zeta + \xi)^2 } \,
   {\rm d}\zeta \, = \,
   {1 \over \xi} - {\rm e}^\xi E_1(\xi),
\nonumber \\
   Q_2^{\rm ph}(\xi,0,0) & = &  \int_0^\infty
   { {\rm e}^{-\zeta g -g\xi+\xi}
   \over \zeta + \xi } \, {\rm d}\zeta
    =  {\rm e}^\xi
   E_1(g \xi).
\label{Q_21}
\end{eqnarray}
The electron transport properties
in the ultra-quantum limit
$\nu \ll 1$ are discussed, for instance, by Yakovlev (1980, 1984).

\section{Energy averaging of relaxation time or $\Psi(E)$} 
The results of Sects. 4.4 and 4.5 fully determine
$\Psi(E)$. The next step is
to evaluate the longitudinal electron transport coefficients
from Eqs.~(\ref{kappa}) and (\ref{Int_Phi}).
This should be done by a numerical energy integration
in Eq.~(\ref{Int_Phi}).
The integration corresponds to statistical energy averaging
of the relaxation time which {\it broadens} the
quantum oscillations. The broadening is
produced by the {\it thermal effects}: all electrons with
energies $| \ep - \mu | \la \kB T$  are known to
contribute into the transport properties.

Actually, quantum oscillations
can also be broadened by other mechanisms which have been
neglected in Sect.~3 and 4
(since we have used
the Born approximation and assumed strictly elastic electron
scattering, see Paper\,I).
Exact theory of the broadening would have been very sophisticated, and
it is still absent.
We incorporate the effects of the broadening
in an approximate manner. In addition
to the thermal broadening, we will take into account
the {\it collisional} broadening of the Landau levels,
and the broadening due to weak {\it inelasticity}
of electron collisions.

To include these effects we replace 
$(\ep-\mu)$ by $(\ep-\mu)T/T_k$ 
in the integrands of Eq.~(\ref{Int_Phi}),
with $T_k = T + [(\gamma + \Delta \ep)/(2 \pi \kB )]$. 
Here $\gamma$ is the collisional width
of the Landau energy levels which 
we set equal
to $\gamma = \hbar / \tau_0$, $\tau_0$ being the
field-free electron relaxation time (Sect.~4.3; Paper\,I).
We include $\gamma$ into $T_k$ in a way
familiar to the semi-quantitative
treatment of magnetic oscillations
in terrestrial metals (e.g., Shoenberg 1984).
In solid matter, $T < T_{\rm m}$, we set
$\gamma = \gamma_{\rm ph} + \gamma_{\rm imp}$
(even for $T<T_U$, ignoring
the failure of our low-$T$ phonon treatment for
very low $T$, Sect. 3.1).
The quantity $\Delta \ep$ is a typical
energy transfer of an electron in a collision event.
We set $ (\Delta \ep)_{\rm ion}
\approx \hbar \omega_{\rm p} + v_{\rm i} p_{\rm F0}$,
$ (\Delta \ep)_{\rm ph} \approx \hbar \omega_{\rm p}G_0(t)$,
and $(\Delta \ep)_{\rm imp} \approx 0$,
where $\omega_{\rm p}$ is the ion plasma frequency,
$v_{\rm i}$ is the ion thermal velocity,
$G_0(t)$ (with $t=T/T_{\rm p}$)
describes reduction of
typical frequencies of phonons excited or absorbed
by electrons at
$T \ll T_{\rm D}$.
We set $G_0(t) \sim G(t)$, where $G(t)$ is the 
reduction factor introduced in Sect.\,3.1. 
Our choice of $\gamma$ and $\Delta \ep$ is rather
phenomenological. However the results
(the transport coefficients) are not too sensitive
to this choice. If more accurate values of $\gamma$ and
$\Delta \ep$ appear in the future, they could
easily be incorporated in our calculation scheme.

Using Eqs.~(\ref{kappa}),
(\ref{Int_Phi}), and (\ref{Psi}), we obtain the following
practical equations for the longitudinal
electric conductivity $\sigma$, thermal conductivity
$\kappa$ and thermopower $\alpha$ as functions of
electron chemical potential $\mu$, temperature $T$, magnetic
field $B$, and nuclear composition of matter:
\begin{eqnarray}
       \sigma & = &  I_0,~~~
       \kappa =  {\pi^2 k_{\rm B}^2 \, T  \over 3 e^2} \,
       J_\kappa,~~~
       \alpha = {\kB \over e} J_\alpha,
\label{Transport_pr} \\
       J_\kappa &=& I_2 - {I_1^2 \over I_0},~~
       J_\alpha = {I_1 \over I_0},
\nonumber \\
       I_j & = & {\sigma_0 \over \theta} \,
          \int_1^\infty \, {\rm d}E \, \Psi(E) \, \zeta^j \,
          { {\rm e}^\zeta \over ({\rm e}^\zeta -1)^2},
\nonumber \\
      \sigma_0^{\rm ion} & = & {m^4 c^6 \over
            4 \pi^3 \hbar^3 e^2 n_{\rm i} Z^2}
        \approx  2.473 \times 10^{22} \,
         {A \over \rho_6 Z^2}~{\rm s}^{-1},
\nonumber \\
      \sigma_0^{\rm imp} & = & \sigma_0^{\rm ion}
        \,{Z^2 \, n_{\rm i} \over (Z_{\rm imp}-Z)^2 \, n_{\rm imp}},
\nonumber \\
     \sigma_0^{\rm ph} & = & {m^2 c^4 \over
        \hbar k_{\rm B} T u_{-2}}
       \approx  1.794 \times 10^{21} \, {1 \over T_7}~{\rm s}^{-1}.
\label{const_pr}
\end{eqnarray}
In this case $\zeta = (\ep - \mu)/ (k_{\rm B}T_k)$,
and $\theta=k_{\rm B}T_k/(mc^2)$.

Equations (\ref{Transport_pr})
allow us to evaluate
the required transport coefficients for any parameters
of study indicated in Sect.~2 and 4.4.

The integration in $I_j$ has been optimized taking into account 
oscillating behaviour of $\Psi(E)$
and the square root singularities at 
the Landau thresholds $E_n=\sqrt{1+2bn}$.
The integrals are presented as sums of
separate integrals over all those intervals $[E_n, E_{n+1}]$
which fall in range from $E= E_\mu - 40 \theta$
to $E=E_\mu + 40 \theta$, with $E_\mu = \mu /(mc^2)$.

For high enough temperatures, when this range
covers more than 25 thresholds of the Landau levels $n$,
integration in $I_j$ over every selected interval $[E_n, E_{n+1}]$
is carried out using Eq.~25.4.34 of Abramowitz \& Stegun (1972)
with 4 mesh points between neighboring Landau thresholds.
This procedure is specifically adopted for integrand
functions which behave as $\sqrt{E-E_n}$ in the vicinity
of $E=E_n$.

For low temperatures, a more accurate numerical
integration is required near $E=E_\mu$. 
For this purpose, we select those intervals
$[E_n, E_{n+1}]$ which span the range from
$E=E_\mu - 8 \theta$ to $E_\mu + 8 \theta$.
One or two of the selected intervals
which include the points $E=E_\mu \pm 8 \theta$ 
are further subdivided by these boundary points. 
We integrate over any such interval 
using the 128-point Simpson formula.
The integration over intervals not spanning 
the selected range is done as in the case 
of high temperatures.

This algorithm ensures fast and accurate
(with error $\la$ 0.1\%) evaluation of the integrals $I_j$.

In order to combine the scattering on low-temperature
phonons and impurities at low $T$ we ignore the breakdown
of the low-$T$ phonon approximation at $T \sim T_U$.
We evaluate separately the integrals $I_j$
($j$=0, 1, 2) in Eqs.~(\ref{const_pr}) due to the
phonon and impurity scatterings at $T<T_{\rm m}$. Then we
calculate the total values of $I_j$ for $T<T_{\rm m}$ as
$I_j^{-1}=(I_j^{\rm ph})^{-1} + (I_j^{\rm imp})^{-1} $,
and determine the transport coefficients from
Eqs.~(\ref{Transport_pr}) and (\ref{const_pr}).

For practical purpose, we should also
calculate the electron number
density $n_{\rm e}$ (which specifies mass
density $\rho = m_u n_{\rm e}(A/Z)$) and its derivative
with respect to the chemical potential
that determines the electron screening wavenumber $q_{\rm e}$
according to Eq.~(\ref{q_i}).
The practical expressions for $n_{\rm e}$ and
$\partial n_{\rm e} / \partial \mu$
are given in Appendix~C of Paper\,I. In these expressions
we will also include the broadening of the Landau levels.
The thermal broadening is naturally implanted 
in these equations just as in Eq.\,(\ref{Int_Phi}). The inelastic
scattering can be important for the kinetic properties
but is inadequate for thermodynamics.
Thus, we incorporate the
collisional broadening by replacing
$ T \to T_{\rm th} = T + [\gamma /(2 \pi \kB )]$ in Eqs.~(C1)
and (C3) of Paper~I.

In practice, one often needs the transport coefficients
as a function of mass density $\rho$ rather than the electron
chemical potential $\mu$. The inverse
dependence of $\mu$ on $\rho$ and $T$ can be found by
iterations from the fitting formula for $n_{\rm e}(\mu)$ 
given in Paper\,I. For example, if the electrons populate 
several Landau levels and $\mu>mc^2$, we use the following 
rapidly converging algorithm. 
The zero-order value of $\mu$ can be set equal to
$\mu_0= c \sqrt{(mc)^2+ p_{\rm F0}^2}$,
the field-free value (\ref{mu_0}) in the limit of strong degeneracy.
At any subsequent iteration step $i=1,2,3,\ldots$, we calculate
$n_{\rm e}^{(i)}$ from the fitting formula with $\mu=\mu_{i-1}$.
Then we determine $\mu_i$ for the next iteration as
$\mu_i=c \sqrt{(mc)^2 +  p_i^2 }$, taking
$p_0=p_{\rm F0}$ and $p_i= p_{i-1}(n_e/n_e^{(i-1)})^{1/3}$. 

\section{Results and discussion}                          
Using the results of Sect.~4 and 5, we have created
a computer code which calculates the longitudinal
electric and thermal conductivities and the thermopower
of degenerate electrons for densities
$10^4$ g~cm$^{-3} < \rho \la 4 \times 10^{11}$ g~cm$^{-3}$,
magnetic fields $B \la 10^{14}$~G, and arbitrary nuclear composition
in the outer crust of a neutron star
(as described in Sects.~2, 3.1, and 4.4). The code is based
on theoretical formalism that is strictly justified for the Coulomb
scattering of electrons on ions at $T>T_{\rm m}$,
for the scattering on high-temperature
phonons at
$T_{\rm D} \la T \la T_{\rm m}$,
and for the scattering on
impurities at $T < T_{\rm m}$. The results are also
valid (semi-quantitatively)
for the low-temperature phonon scattering
at $T_U \la T \ll T_{\rm D}$ as discussed in Sect.~3.1.

Figures \ref{fig6.1} -- \ref{fig6.4} show
typical density dependence of the longitudinal transport coefficients
in a neutron star crust with a strong magnetic field.

%
     \begin{figure*}
 \begin{center}
 \leavevmode
 \epsfysize=100mm 
 \epsfbox{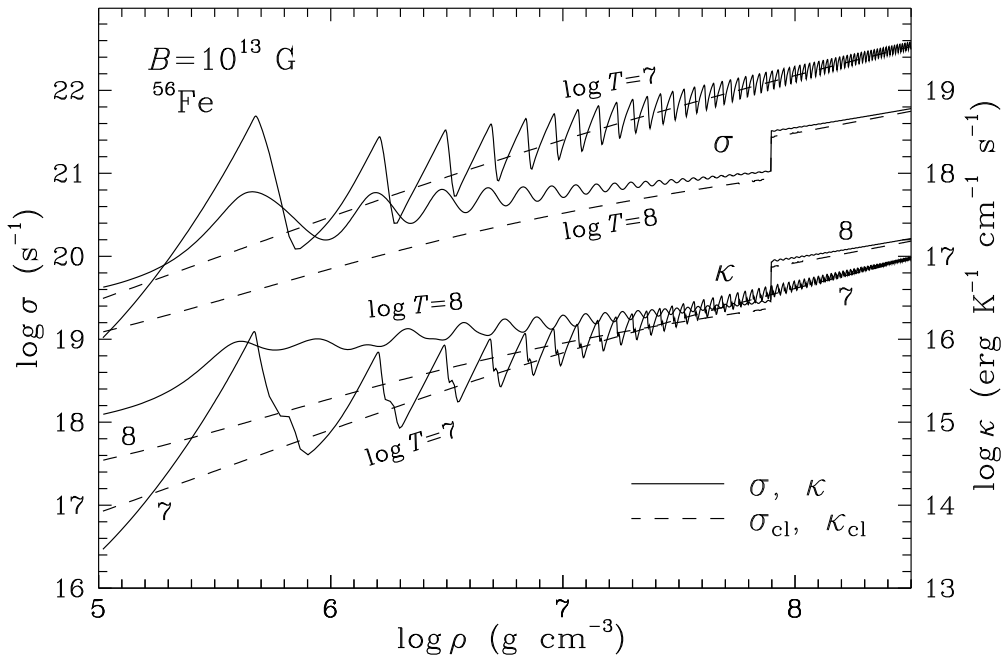}
 \end{center}
     \caption[ ]{
   Electric (left vertical scale, upper lines) and thermal
   (right scale, lower lines) conductivities
   of $^{56}$Fe matter vs density for $B=10^{13}$~G (solid lines)
   and $B=0$ (dashes) at $T=10^7$~K and $10^8$~K.
}
\label{fig6.1}
\end{figure*}
%
%
\begin{figure*}
 \begin{center}
 \leavevmode
 \epsfysize=100mm 
 \epsfbox{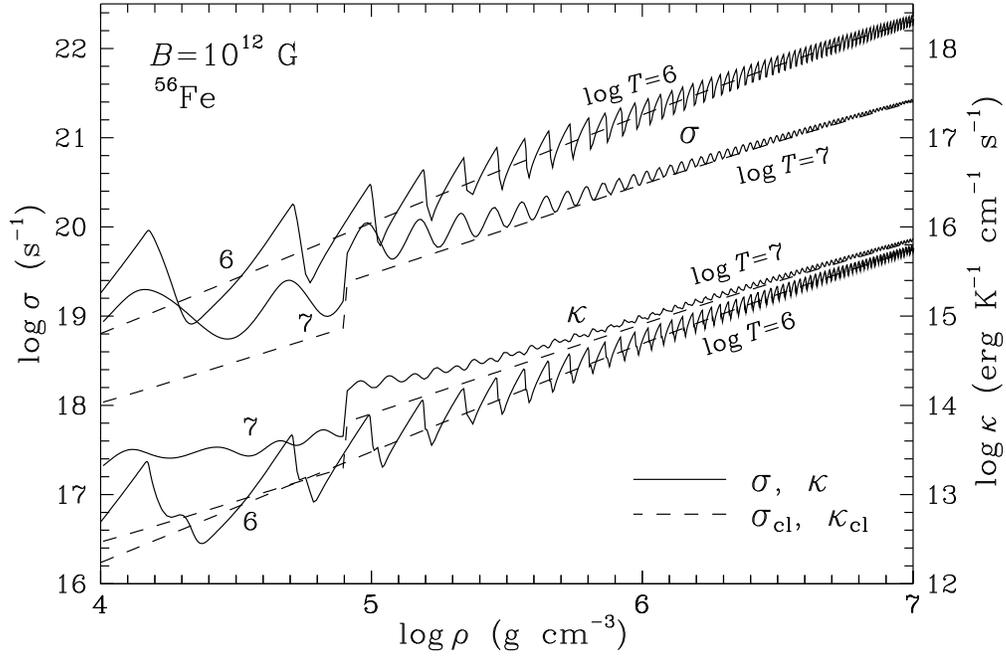}
 \end{center}
\caption[ ]{
   Same as in Fig.\ \ref{fig6.1} for $B=10^{12}$~G
   at $T=10^6$ and $10^7$~K.
}
\label{fig6.1a}
\end{figure*}
%
%
\begin{figure*}
 \begin{center}
 \leavevmode
 \epsfysize=100mm 
 \epsfbox{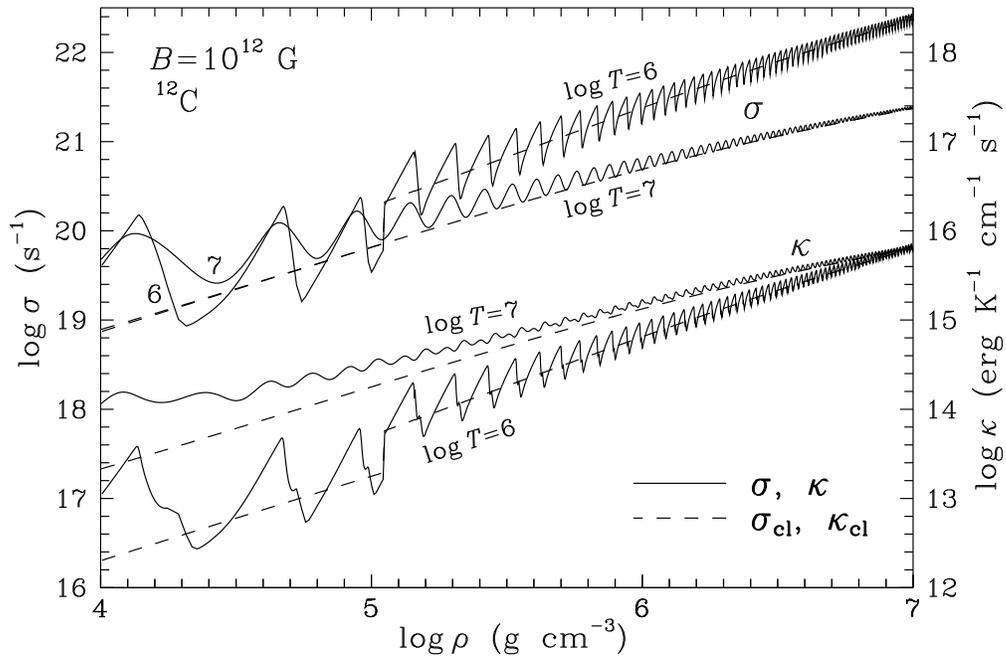}
 \end{center}
\caption[ ]{
   Same as in Fig.\ \ref{fig6.1a} for $^{12}$C matter.
}
\label{fig6.2}
\end{figure*}
%
%
\begin{figure*}
 \begin{center}
 \leavevmode
 \epsfysize=100mm 
 \epsfbox{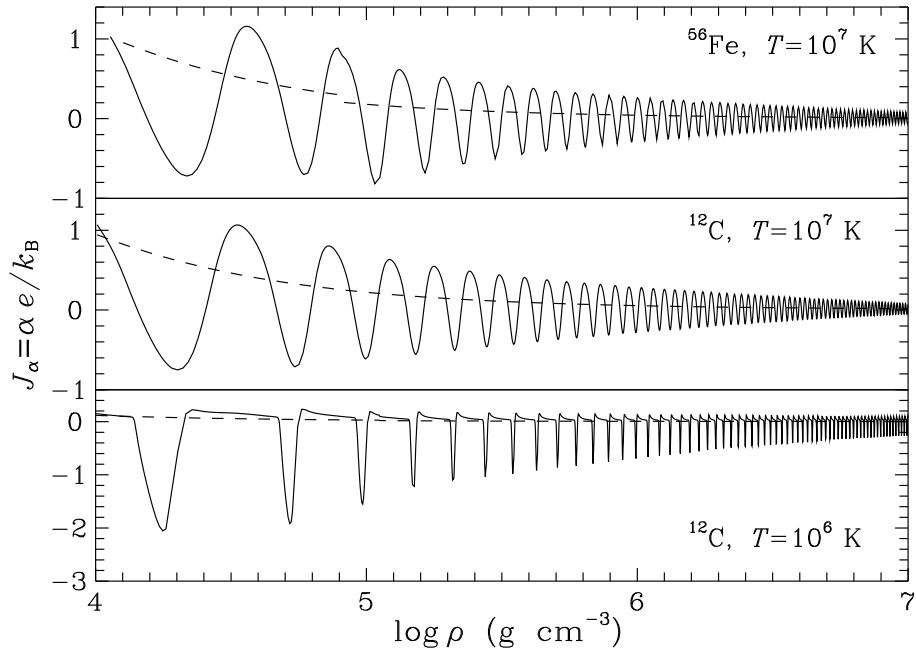}
 \end{center}
\caption[ ]{
   Dimensionless longitudinal thermopower for
   $^{56}$Fe and $^{12}$C matter,
   $B=10^{12}$~G and $T=10^6$ and 10$^7$~K. Dashes show
   the $B=0$ curves.
}
\label{fig6.3}
\end{figure*}
%
%
\begin{figure*}
 \begin{center}
 \leavevmode
 \epsfysize=100mm 
 \epsfbox{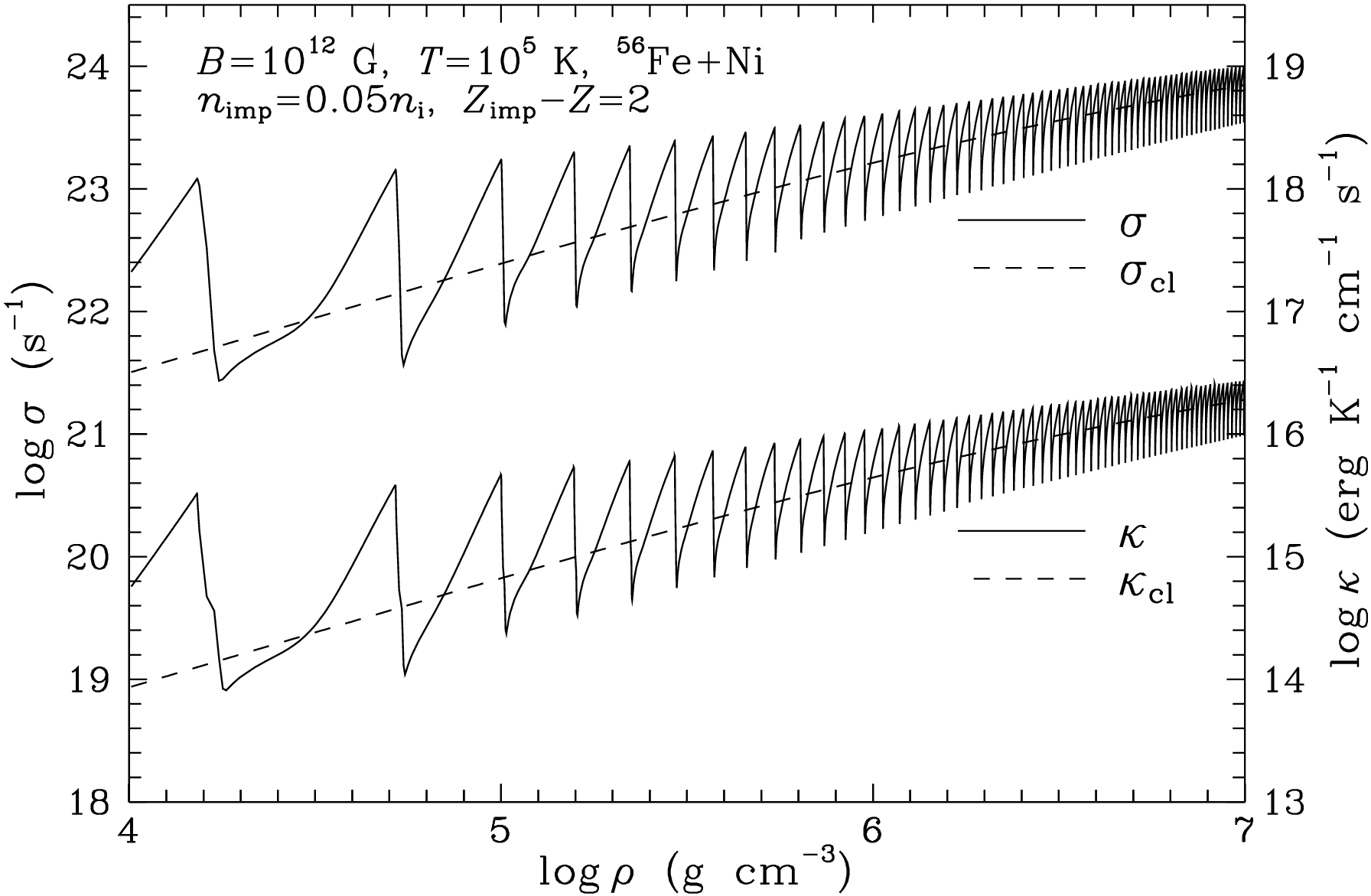}
 \end{center}
\caption[ ]{
   Same as in Figs.~\ref{fig6.1} -- \ref{fig6.2}
   for electron scattering on charged impurities in
   $^{56}$Fe matter at $B=10^{12}$~G.
}
\label{fig6.4}
\end{figure*}

Figure \ref{fig6.1} demonstrates quantum oscillations of
the electric and thermal conductivities with increasing density
in $^{56}$Fe matter for $B=10^{13}$~G
at two temperatures, $T=10^7$~K and $10^8$~K. Every oscillation
is associated with population of a new Landau level,
starting from the first excited level $n=1$ at
$\rho \approx \rho_B \approx 5 \times 10^5$ g~cm$^{-3}$
(cf. Fig.~1). In the displayed density
range ($10^5$ g~cm$^{-3} < \rho < 3 \times 10^8$ g~cm$^{-3}$),
up to about 60 Landau levels are populated.
For comparison, dashed lines show the
conductivities $\sigma_{\rm cl}$ and
$\kappa_{\rm cl}$ in the absence of the magnetic field.

For the lower temperature, $T=10^7$ we have
$T \ll T_B$ (see Fig.~1).
The adopted magnetic field
$B=10^{13}$~G is
quantizing (Sect.~2), and
the oscillations are quite pronounced. At this temperature,
matter is solid ($T<T_{\rm m}$), and the oscillations
are produced by the scattering of electrons on 
phonons. These oscillations are significantly amplified
by the Debye--Waller suppression of the electron scattering
(Paper\,I). In the domain of strongly
quantizing magnetic field ($\rho \la \rho_B$, Sect.~2),
the transport coefficients differ drastically from those
at $B=0$. With increasing $\rho$ in the domain of weakly
quantizing fields ($\rho \gg \rho_B$, $T \ll T_B$),
the oscillations become weaker, and the conductivities are
seen to be close to the non-magnetic ones.
Every oscillation of the electric conductivity contains one
maximum and subsequent dip, while the oscillation of the thermal
conductivity may show a secondary maximum as explained, for
instance, by Yakovlev (1980, 1984). The amplitudes of quantum
oscillations of the electron conductivity are always stronger
than those of the thermal conductivity.

For the higher temperature $T=10^8$~K in Fig.\ \ref{fig6.1},
the magnetic field
$B=10^{13}$~G is less quantizing (the ratio $T_B/T$ remains
larger than unity but it is 10 times smaller
than for $T=10^7$~K). Accordingly, the quantum oscillations
are noticeably weaker due to the thermal broadening of the
Landau levels. The broadening is more
pronounced in the thermal conductivity than in the
electric conductivity.
The jumps of the transport coefficients at $\rho \approx 10^8$
g~cm$^{-3}$ are associated with solidification. For
lower $\rho$ (at given $T$), matter is melted, and the
electron relaxation is produced by the Coulomb scattering on
ions. For higher $\rho$, ions solidify, and the electrons
scatter on phonons just as at $T=10^7$~K.
At higher $T \sim T_B$, the oscillations would be entirely
smeared out by the thermal broadening.
Note that the transport coefficients
in the ion liquid with the magnetic field at $T=10^8$~K
are systematically larger
than those in the non-magnetic matter.
This means that the transport coefficients
averaged over magnetic
oscillations are larger than the coefficients for $B=0$.
This difference between the oscillation-averaged and
non-magnetic coefficients is explained by the classical effect
of the electron Larmor rotation on the Coulomb logarithm
in the transport
cross section for the electron--ion scattering: the non-quantizing
magnetic field decreases the Coulomb logarithm
(and amplifies the electric and thermal conductivities)
by reducing the maximum effective impact parameter
(e.g., Yakovlev 1980).

Figure \ref{fig6.1a} shows quantum oscillations
of the transport coefficients in
iron matter with weaker magnetic field, $B=10^{12}$~G,
at lower $T$ (10$^6$ and 10$^7$~K). Since the magnetic
field is lower, the oscillations
associated with population of the same Landau levels
occur at lower $\rho$ (cf. Fig.~1). If $T=10^6$~K,
matter is solid in the displayed density range,
$10^4$~g~cm$^{-3} < \rho < 10^7$ g~cm$^{-3}$,
while for $T=10^7$~K it solidifies at $\rho \approx
8 \times 10^4$ g~cm$^{-3}$. Qualitative behavior of
the curves is the same as in Fig.~\ref{fig6.1}.

Figure \ref{fig6.2} presents quantum oscillations of
the thermal and electric conductivities
in carbon matter for $B=10^{12}$~G
at $T=10^6$ and $10^7$~K. The main features are the same
as in Figs.~\ref{fig6.1} and \ref{fig6.1a}, although
the melting temperature
$T_{\rm m}$ ($\Gamma = 172$,
Eq.~(\ref{Gamma})) is lower than for iron.
Population of the Landau levels is mainly independent
of ion species ($A,Z$), and occurs at
the same $\rho$ as in Fig.~\ref{fig6.1a}.
If $T=10^7$~K, matter is
melted, in the displayed parameter range.
In the case of $T=10^6$~K, matter is liquid for
$\rho \la 10^5$ g~cm$^{-3}$ and solid at higher $\rho$.
In this case quantum oscillations are pronounced stronger 
than at $T=10^7$~K.

Figure \ref{fig6.3} shows quantum oscillations of the
dimensionless longitudinal thermopower $J_\alpha = \alpha e/\kB$
(defined by Eq. (\ref{Transport_pr}))
for $B=10^{12}$~G. The upper panel
corresponds to iron matter at $T=10^7$~K (cf. Fig.~\ref{fig6.1a}).
Two lower panels correspond
to carbon matter for the same conditions as in Fig.~\ref{fig6.2}.
The thermopower is known to be more sensitive to the
electron scattering mechanism, than the conductivities.
Its quantum oscillations are seen to be stronger, and
more complicated. The oscillating thermopower differs significantly
from that at $B=0$, and it can change sign
(Yakovlev 1980, 1984). The jumps associated with the solidification
are pronounced weaker than in the conductivities.

The curves displayed in Figs.~\ref{fig6.1} -- \ref{fig6.3}
are calculated taking into account 
collisional and inelastic-scattering 
broadening of the Landau levels (Sect.~5). 
These broadening mechanisms
appear to be less significant
than the thermal broadenings 
for the scattering on ions in
Coulomb liquid and on phonons in the
crystalline matter. Nevertheless non-thermal broadenings become more
important with decreasing temperature. In particular,
the inelastic-scattering broadening becomes significant
at $T  \ll
T_{\rm D}$, when the employed approximation of
high-temperature phonons breaks (Sect.~3.1).
The collisional and inelastic-scattering broadenings
also become important with decreasing density
near the domain of incomplete ionization (Sect.~2).

Figure~\ref{fig6.4}
shows oscillations
of the electric and thermal conductivities
produced by electron scattering on charged impurities
in iron matter with $B=10^{12}$~G
at low temperature $T=10^5$~K
assuming, for illustration, that the scattering
on phonons is inefficient.
The impurity number
density is supposed to be $n_{\rm imp} = 0.05 n_{\rm i}$,
and the impurity charge number is $Z_{\rm imp}$=28.
The impurity screening wavenumber $q_{\rm imp}$ in Eq.~(\ref{U_imp})
is set equal to $q_{\rm imp} = (4 \pi n_{\rm imp}/3)^{1/3}$.
At this low temperature,
the collisional broadening of the Landau
levels becomes more important.
With decreasing $T$, the thermal broadening would die out,
and the shapes of the oscillations would be entirely determined
by the collisional broadening. Note that we include
the non-thermal broadening mechanisms
in an approximate manner. This introduces an uncertainty
into the transport coefficients at low temperatures.
However the case of low temperatures is not very important
for applications. Anyway, the number density and charge
number of impurities are not known, and this introduces
much larger uncertainty in our knowledge of the transport
properties of very cold stellar matter.

\section{Summary}                                         
We have obtained practical formulae for evaluation
of the longitudinal electric and thermal conductivities and
longitudinal thermopower in degenerate dense matter
($10^4$ g~cm$^{-3} \la \rho \la 4 \times 10^{11}$ g~cm$^{-3}$)
of outer neutron star crusts with strong magnetic fields $B=
10^{10}$ -- $10^{14}$~G (at stronger fields,
the splitting of electron Landau levels concerned with
anomalous electron magnetic moment should be taken into account).
The results are expressed in terms of the
energy averaged electron relaxation time $\tau(\ep)$ 
or the function $\Psi(E)$.
We have found (Sect.~4) accurate and simple analytic fits for $\Psi(E)$,
which are valid in wide ranges of the parameters of stellar
matter for three electron scattering mechanisms: for
the Coulomb scattering on ions in gaseous or liquid phases,
for the scattering on high-temperature phonons in solid
matter, and for the scattering on charged impurities in solids
at low temperatures. We have proposed (Sect.~5) an efficient
energy averaging procedure which allows us to evaluate rapidly
any longitudinal electron transport coefficient. The numerical
examples are given in Sect.~6.

Note that our formalism can easily be extended to higher
densities, $\rho > 4 \times 10^{11}$ g~cm$^{-3}$,
in the inner crust of a neutron star, where free neutrons
appear in matter in addition to electrons and atomic nuclei.
In this case nuclei can occupy substantial part of volume,
and one should take into account finite
sizes of nuclei by multiplying the Fourier image of
the scattering potential, $|U(\vect{q})|^2$ (Sect. 3.1),
by a squared nuclear formfactor (e.g., Itoh et al. 1984).
However, this effect can be shown
to be not very strong. For instance, in the absence of
the magnetic field, it reduces the effective electron collision
frequencies typically by about 20 \%,
in the inner crusts of neutron stars.
Accordingly, 
the present theoretical framework is 
expected to be sufficiently accurate in the inner crusts 
as well.

The results of this work are required for studying
various processes in neutron star crusts
(e.g., Yakovlev \& Kaminker 1994). First of all,
we mention thermal evolution (cooling) of neutron stars.
The outer crust produces thermal isolation of the
stellar interior. The electron thermal conductivity
is most important for calculating the relationship
between the surface and interior temperatures of the star
and for evaluating the distribution of the temperature
over the neutron star surface. The latter distribution
can be strongly anisotropic owing to anisotropic character
of the thermal conductivity. The anisotropy of the
surface temperature leads to modulation of the surface
thermal radiation due to stellar rotation.
The modulation has been observed with {\it ROSAT} in soft X-ray
radiation of several neutron
stars, in particular, PSR 0656+14
(Finley et al.\ 1992, Anderson et al.\ 1993) and Geminga
(Halpern \& Holt 1992, Halpern \& Ruderman 1993).
Correct interpretation of observations requires
the knowledge of the distribution of the effective temperature
over the neutron star surface combined with the solution of
the radiative transfer problem in the neutron star atmosphere
(e.g., Pavlov et al.\ 1995). Much work has already been done
by Hernquist (1984),
Van Riper (1988, 1991) and Schaaf (1988, 1990)
in calculating the temperature profiles in neutron star
crusts with quantizing magnetic fields, and in analysing
the cooling of magnetized neutron stars.
Many important and reliable results concerning the temperature profiles
have already been obtained (particularly, by
Van Riper 1988) in the
approximation of uniform plane-parallel layer with
the magnetic field normal to the surface.
However the overall problem of the surface temperature
distribution is complicated and requires further study
(as discussed, e.g.,
by Yakovlev \& Kaminker 1994). The solution of this problem
can be based, to some extent, on the above results.
Secondly, the electric conductivity
and thermopower are important for understanding the evolution
of neutron star magnetic fields
(e.g., Urpin et al.\ 1986, 1994). We plan to
construct the models of outer crusts of neutron stars
with strong magnetic fields in our subsequent works.

The Fortran computer code for evaluating the
transport properties based on the above results
is distributed freely by electronic mail
upon request.

\begin{acknowledgements} 
Useful discussions with C.J.\,Pethick 
and valuable comments of K.A.\,Van Riper 
are gratefully acknowledged. 
This work was supported in part by RBRF (grant 96-02-16870),
ISF (grant R6A300), and INTAS (grant 94-3834). 
A.Y.P.\ is grateful to the Nordita staff and especially
to C.J.\,Pethick for hospitality.
\end{acknowledgements}

%

\end{document}